\documentclass[12pt]{article}

\begin{document}

\title{\textbf{Kinematic Self-Similar Heat Conducting and Charge Solutions}}

\author{M. Sharif \thanks{msharif@math.pu.edu.pk} and Wajiha Javed \\
\\Department of Mathematics, University of the Punjab,\\Quaid-e-Azam
Campus Lahore-54590, Pakistan.}
\date{}

\maketitle
\begin{abstract}
The objective of this paper is to study the plane symmetric
kinematic self-similar heat conducting fluid and charge dust
solutions of the Einstein field equations. These solutions are
classified according to self-similarity of the first, second, zeroth
and infinite kinds with different equations of state. We take the
self-similar vector to be tilted, orthogonal and parallel to the
fluid flow.  For heat conducting fluid, it is found that there exist
only \emph{one} solution in parallel case. In all other
possibilities, these solutions reduce to the perfect fluid kinematic
self-similar solutions. For charge dust case, we also obtain only
\emph{one} kinematic self-similar solution.
\end{abstract}


\section{Introduction}

General Relativity (GR) demonstrates the most dominant attribute of
the universe, gravity, in terms of geometry of the curved
spacetimes. The Einstein field equations (EFEs) are the core of GR
that are highly non-linear, second order coupled partial
differential equations (PDEs). Due to the mathematical complexity of
these equations, there does not exist any general solution. However,
one can find particular solutions (exact solutions) by imposing
certain symmetry assumptions on the concerned system.
Self-similarity (SS) is one of the techniques which is very helpful
in simplifying the field equations by reducing the number of
variables.

Self-similarity is a scale invariant property which does not change
the solution of the EFEs under any scale transformation. These scale
invariant solutions of the field equations are called self-similar
solutions (SSS). There are many applications of such solutions in
astrophysics and cosmology. These solutions are used in the
discussion of extreme physical complications. The astrophysical
applications include gravitational collapse and the occurrence of
naked singularities while the cosmological applications include
features of gravitational clustering and cosmic voids.

The concept of SS in GR was defined for the first time by Cahill and
Taub \cite{KSPSS8}, who discussed SS of the first kind by the
existence of homothetic vector field in the spacetime. Carter and
Henriksen \cite{KSPSS9} defined SS of the second, zeroth and
infinite kinds. Maeda et al. \cite{KSPSS3} worked on kinematic
self-similar (KSS) perfect fluid and dust solutions of spherically
symmetric spacetime for the tilted, parallel and orthogonal cases.

Carr et al. \cite{KSPSS16} described physical features of
spherically symmetric self-similar perfect fluid models with EOS
$p=k\rho$. Further, they explored the KSS vector associated with
critical behavior observed in gravitational collapse. Coley and
Golaith \cite{KSPSS17} investigated self-similar spherically
symmetric cosmological models with a perfect fluid and a scalar
field with an exponential potential. Sintes et al. \cite{KSPSS14}
considered plane, spherical and hyperbolical symmetric spacetimes to
discuss the KSS perfect fluid solutions of the infinite kind.

The symmetries of plane symmetric spacetime has been considered by
using different procedures \cite{MS}. Sharif and Sehar
\cite{KSPSS25}-\cite{KSPSS7} investigated the KSS solutions for
plane and cylindrical symmetric spacetimes. The analysis has been
given for perfect fluid and dust cases with tilted, parallel and
orthogonal vectors. The same authors \cite{KSPSS18} also discussed
the physical properties of homothetic solutions for spherically,
cylindrically and plane symmetric spacetimes. Sharif with his
collaborators {\cite{SAJ TH,JAV TH}} found the KSS solutions of the
most general cylindrically symmetric spacetimes for perfect and dust
fluid.

Recently, Misthry et al. \cite{KSPSS24} studied radiative collapse
of the realistic models (radiating stars) by using non-viscous
heat conducting fluid. Herrera \cite{KSPSS25c} found that heat
produced inertia in the dissipative collapse. Nath et al.
\cite{KSPSS25b} have studied the gravitational collapse of heat
conducting non-viscous fluid. They have found that electromagnetic
field reduces the pressure at the boundary which is balanced by
the heat flux.

In this paper, we explore the influence of heat flux and
electromagnetism on KSS solutions. For this purpose, we study
non-viscous heat conducting fluid and charge dust as a gravitating
material. The paper has been organized as follows. In section 2, the
EFEs are simplified by taking KSS vector of the first, second,
zeroth and infinite kinds and the resulting system of ODEs is solved
analytically for the plane symmetric spacetimes. We take non-viscous
heat conducting fluid and dust fluid, when the KSS vector is tilted,
orthogonal and parallel to the fluid flow. Sections 3 is devoted to
study the KSS solutions by taking charge dust fluid. The last
section gives an outlook of the results.

\section{Kinematic Self-similar Heat Conducting Fluid Solutions}

The most general plane symmetric spacetime is given in the form
\cite{BK16STPH1}
\begin{equation}\label{a:1}
ds^{2}=e^{2\nu(t,x)}dt^{2}-e^{2\mu(t,x)}dx^2
-e^{2\lambda(t,x)}(dy^2+dz^2),
\end{equation}
where $\nu,~\mu$ and $\lambda$ are arbitrary functions of $t$ and
$x$. This metric has three isometries given as
$\xi_1=\partial_y,~\xi_2=\partial_z,~\xi_3=y\partial_z-z\partial_y$.
In plane symmetric spacetime, the most general spacetime (\ref{a:1})
can be simplified to the following form
\begin{equation}\label{a:2}
ds^2=e^{2\mu(t,x)}(dt^2-dx^2)-e^{2\lambda(t,x)}(dy^2+ dz^2).
\end{equation}
The energy-momentum tensor for a non-viscous heat conducting fluid
is given by \cite{BK16STPH}
\begin{equation}\label{a:3}
T_{ab}=[\rho(t,x)+p(t,x)]u_{a}u_{b}-
p(t,x)g_{ab}+q_{a}u_b+q_bu_a,\quad(a,b=0, 1, 2, 3)
\end{equation}
where $\rho$ and $p$ are density and pressure respectively,
$u_{a}=(e^{\mu}, 0, 0, 0)$ is the four-velocity of the fluid
element in the co-moving coordinate system and
$q^{a}=(0,q(t,x),0,0)$ is the heat flow vector. Notice that
$q^{a}u_{a}=0$ for $x$ directed heat flow. The EFEs for the line
element (\ref{a:2}) can be written as
\begin{eqnarray}\label{a:4}
\kappa\rho&=& e^{-2\mu}(-3{\lambda_{x}}^{2}-2\lambda_{xx}
+{\lambda_{t}}^{2}+2{\lambda_{x}{\mu_{x}}}+2{\lambda_{t}}{\mu_{t}}),\\
\label{a:5}\kappa q
&=&2e^{{-3}\mu}(\lambda_{tx}-\lambda_{t}\mu_{x}+\lambda_{t}\lambda_{x}-\mu_{t}\lambda_{x}),\\
\label{a:6} \kappa p &=&
e^{-2\mu}({\lambda_{x}}^{2}+2\lambda_{x}\mu_{x}-2\lambda_{tt}+2\lambda_{t}\mu_{t}-3{\lambda_{t}}^{2}),\\
\label{a:7} \kappa p &=& e^{-2\mu}(\mu_{xx}+{\lambda_{x}}^{2}
+\lambda_{xx}-\lambda_{tt}-{\lambda_{t}}^{2}-\mu_{tt}).
\end{eqnarray}
The conservation of energy-momentum tensor, $T^{ab}_{~~;b}=0,$
gives
\begin{eqnarray}\label{a:8}
{\mu_{t}}&=&-\frac{\rho_{t}}{(\rho+p)}-2\lambda_{t}-
\frac{e^{\mu}}{(\rho+p)}{({q_x}+3q{\mu_x}+2q{\lambda_x})},\\
\label{a:9}
\mu_{x}&=&-\frac{p_{x}}{(\rho+p)}-\frac{1}{(\rho+p)}{e^{\mu}}
{(q_{t}+2q{\lambda_t}+3{\mu_t}q)}.
\end{eqnarray}
The vector field $\underline{\xi}$ for a plane symmetric spacetime
is
\begin{equation}\label{a:10}
\xi^{a}\frac{\partial}{\partial
x^{a}}=h_{1}(t,x)\frac{\partial}{\partial
t}+h_{2}(t,x)\frac{\partial}{\partial x},
\end{equation}
where $h_{1}$ and $h_{2}$ are arbitrary functions of $t$ and $x$.
For $h_{2} =0$, this gives parallel case while for $h_{1}=0$, we
have the orthogonal case. When both $h_{1}$ and $h_{2}$ are
non-zero we have the most general case known as the tilted case.

A KSS vector $\underline{\xi}$ satisfies the following conditions
\begin{eqnarray}\label{A:A}
\pounds_{\underline{\xi}}h_{ab}=2\delta
h_{ab},\quad\pounds_{\underline{\xi}}u_a=\alpha u_a,
\end{eqnarray}
where $\alpha$ and $\delta$ are dimensionless constants and
$h_{ab}=g_{ab}-u_a u_b$ is the projection tensor. The ratio,
$\frac{\alpha}{\delta}$, is called the similarity index which
gives rise to the following two cases:
\begin{eqnarray*}
1.\quad \delta \neq 0, \quad 2.\quad \delta =0.
\end{eqnarray*}
It is mentioned here that self-similar variables and the metric
functions for all kinds of tilted, orthogonal and parallel flow
remain the same as found earlier \cite{KSPSS21}. The first case
gives self-similarity of the first, zeroth and second kinds while
the second case gives infinite kind. We would like to omit the
details here as it is given extensively in literature
\cite{KSPSS21}. It is mentioned here that there does not exist
self-similar variable and the corresponding metric functions of the
tilted infinite kind for the line element (\ref{a:2}). Also, these
quantities do not exist for the parallel (except for the second and
infinite kinds) and orthogonal (except for the zeroth kind) cases.

In this paper, we focus on the following two kinds of polytropic
equations of state (EOS) \cite{KSS NET}. The first EOS, denoted by
{EOS(1), is $p=k\rho^\gamma$, where $k$ and $\gamma$ are
constants. We assume that $k\neq0$ and $\gamma\neq0,1$. Another
EOS, denoted by EOS(2), is $p=k
n^\gamma,~\rho=m_{b}n+\frac{p}{\gamma-1}$, where the constants
$m_b$ and $n(t,x)$ correspond to the baryon mass and baryon number
density respectively. Here we assume that $k\neq0$ and
$\gamma\neq0,1$. We also treat an EOS, i.e., EOS(3) by $p=k\rho$,
where we assume that $-1\leq k \leq1$ and $k\neq0$.

\subsection{Tilted Fluid Case}

\subsubsection{Self-similarity of the First Kind}

It follows from the EFEs that the energy density $\rho$, pressure
$p$ and heat flux $q$ must take the following form
\begin{eqnarray}
\kappa\rho(t,x)&=& \frac{1}{x^2}\rho(\xi),\label{b:2}\\
\kappa q(t,x)&=&\frac{1}{xt}q(\xi),\label{b:3}\\
\kappa p(t,x)&=& \frac{1}{x^2}p(\xi),\label{b:4}
\end{eqnarray}
where the self-similar variable $\xi$ is $\frac{x}{t}$. When the
EFEs and the equations of motion for the matter field are
satisfied, it yields a set of ODEs and hence
Eqs.(\ref{a:4})-(\ref{a:9}) reduce to
\begin{eqnarray}
\dot{\rho} &=&-(\dot{\mu}+2\dot{\lambda})(\rho+p)
+e^\mu(\dot{q}+q+3q \dot{\mu}+2q\dot{\lambda}),\label{b:6:1}\\
2p-\dot{p} &=& \dot{\mu}(\rho+p),\label{b:6:2}\\
{e^{2\mu}}\rho &=&
-4\dot{\lambda}-3{\dot{\lambda}}^{2}-2\ddot{\lambda}-1+2\dot{\mu}+2\dot{\lambda}\dot{\mu},\label{b:6:3}\\
0&=& {\dot{\lambda}}^{2}+2\dot{\lambda}\dot{\mu},\label{b:6:4}\\
e^{3\mu}
q&=&-2(\ddot{\lambda}+{\dot{\lambda}}^2+\dot{\lambda}-\dot{\mu}-2\dot{\lambda}\dot{\mu}),\label{b:6:5}\\
{e^{2\mu}}p&=&1+2\dot{\lambda}+{\dot{\lambda}}^2
+2\dot{\mu}+2\dot{\lambda}\dot{\mu},\label{b:6:6}\\
0&=& 2\dot{\lambda}\dot{\mu}-2\ddot{\lambda}
-3{\dot{\lambda}}^2-2\dot{\lambda},\label{b:6:7}\\
{e^{2\mu}}p &=& \ddot{\lambda}+{\dot{\lambda}}^2 +\dot{\lambda}+\ddot{\mu}-{\dot{\mu}},\label{b:6:8}\\
0&=&-\ddot{\lambda}-{\dot{\lambda}}^2-\dot{\lambda}-\ddot{\mu}-{\dot{\mu}},\label{b:6:9}\\
0&=&\dot{q}+q+2q\dot{\lambda}+3\dot{\mu}q,\label{b:6:10}
\end{eqnarray}
where dot ($.$) represents derivative with respect to $\ln\xi$.
Equation (\ref{b:6:4}) gives either $\dot{\lambda}=0$ or
$\dot{\lambda}=-2\dot{\mu}$. For $\dot{\lambda}=0$,
Eqs.(\ref{b:6:6})-(\ref{b:6:9}) give contradiction. For
$\dot{\lambda}=-2\dot{\mu}$, Eqs.(\ref{b:6:6})-(\ref{b:6:8}) yield
contradiction.

\subsubsection{Self-similarity of the Second Kind}

The EFEs imply that $\rho,~p$ and $q$ in terms of $\xi$, i.e.,
\begin{eqnarray}
\label{b:10}\kappa\rho(t,x) &=& \frac{1}{x^2}[\rho_1(\xi)+\frac{x^2}{t^2}\rho_2(\xi)],\\
\label{b:11}\kappa q(t,x)&=&\frac{1}{xt}q(\xi),\\
\label{b:12}\kappa
p(t,x)&=&\frac{1}{x^2}[p_1(\xi)+\frac{x^2}{t^2}p_2(\xi)],
\end{eqnarray}
where $\xi=\frac{x}{(\alpha t)^\frac{1}{\alpha}}$. A set of ODEs is
obtained when the EFEs and the equations of motion for the matter
field are satisfied for the $O[(\frac{x}{t})^0]$ and
$O[(\frac{x}{t})^2]$ terms separately. Equations
(\ref{a:4})-(\ref{a:9}) reduce to the following
\begin{eqnarray}
\label{b:14}\dot{\rho_1}
&=&-(\dot{\mu}+2\dot{\lambda})(\rho_1+p_1)+{e^\mu}\alpha(\dot{q}+q+3q\dot{\mu}+2q\dot{\lambda}),\\
\label{b:15}\dot{\rho_2}+2\alpha\rho_2&=&-(\dot{\mu}+2\dot{\lambda})(\rho_2+p_2),\\
\label{b:16}-\dot{p_1}+2p_1&=&\dot{\mu}(\rho_1+p_1),\\
\label{b:17}-\dot{p_2}&=&\dot{\mu}(\rho_2+p_2)+\frac{e^{\mu}}{\alpha}{(-\dot{q}-\alpha
q -2q\dot{\lambda}-3\dot{\mu}q)},\\
\label{b:18}{e^{2\mu}}\rho_1&=&-4\dot{\lambda}-3{\dot{\lambda}}^{2}-2\ddot{\lambda}-1
+2\dot{\mu}+2\dot{\mu}\dot{\lambda},\\
\label{b:19}\alpha^2e^{2\mu}\rho_2 &=&{\dot{\lambda}}^{2}+2\dot{\mu}\dot{\lambda},\\
\label{b:20}{e^{3\mu}}\alpha
q&=&-2(\ddot{\lambda}+{\dot{\lambda}}^2
+\dot{\lambda}-2\dot{\lambda}\dot{\mu}-\dot{\mu}),\\
\label{b:21}{e^{2\mu}}p_1&=&1+2\dot{\lambda}+{\dot{\lambda}}^2+2\dot{\mu}+2\dot{\lambda}\dot{\mu},\\
\label{b:22}\alpha^2e^{2\mu}p_2&=&-2\ddot{\lambda}-3{\dot{\lambda}}^2
-2\alpha\dot{\lambda}+2\dot{\lambda}\dot{\mu},\\
\label{b:23}{e^{2\mu}}p_1&=&\ddot{\lambda}+{\dot{\lambda}}^2
+\dot{\lambda}+\ddot{\mu}-\dot{\mu},\\
\label{b:24}\alpha^2e^{2\mu}p_2&=&-\ddot{\lambda}-{\dot{\lambda}}^2
-\alpha\dot{\lambda}-\ddot{\mu}-\alpha\dot{\mu}.
\end{eqnarray}

\subsubsection*{EOS (1) and (2)}

If a fluid obeys EOS(1) for $k\neq0$ and $\gamma\neq0,1$, we find
from Eqs.(\ref{b:10}) and (\ref{b:12}) that
\begin{equation}\label{b:25}
\alpha=\gamma,\quad p_1=0=\rho_2,\quad p_2= \frac{k}{(8\pi
G)^{(\gamma-1)}\gamma^2}\xi^{-2\gamma}{\rho_1}^\gamma,
\quad[Case~I]
\end{equation}
or
\begin{equation}\label{b:26}
\alpha=\frac{1}{\gamma},\quad p_2=0=\rho_1,\quad
p_1=\frac{k}{(8\pi
G)^{(\gamma-1)}\gamma^{2\gamma}}\xi^{2}{\rho_2}^\gamma.\quad[Case~II]
\end{equation}
A fluid satisfying EOS(2) for $k\neq0$ and $\gamma\neq0,1$, then
it follows from Eqs.(\ref{b:10}) and (\ref{b:12}) that
\begin{equation}\label{b:27}
\alpha=\gamma,\quad p_1=0,\quad p_2=\frac{k}{{m_b}^{\gamma}(8\pi
G)^{(\gamma-1)}\gamma^2}\xi^{-2\gamma}{\rho_1}^\gamma=(\gamma-1)\rho_2,
\quad [Case~III]
\end{equation}
or
\begin{equation}\label{b:28}
\alpha=\frac{1}{\gamma}, \quad p_2=0, \quad
p_1=\frac{k}{{m_b}^{\gamma}(8\pi
G)^{(\gamma-1)}\gamma^{2\gamma}}\xi^{2}{\rho_2}^\gamma=(\gamma-1)\rho_1.
\quad [Case~IV]
\end{equation}
For the \textbf{Case I}, Eq.(\ref{b:16}) yields either $\dot{\mu}=0$
or ${\rho_1}=0$. In both possibilities, we obtain contradiction. For
the \textbf{Case II}, the general case could not be solved while the
special choice leads to a contradiction. In the \textbf{Case III},
Eq.(\ref{b:16}) yields either $\dot{\mu}=0$ or ${\rho_1}=0$. If
$\dot{\mu}=0$, Eq.(\ref{b:21}) implies that $\dot{\lambda}=-1$,
Eqs.(\ref{b:22}) and (\ref{b:24}) provide $\alpha=2$, while
Eq.(\ref{b:18}) gives ${\rho_1}=0$. If $\gamma=2$, then
Eq.(\ref{b:15}) implies that
$\rho_2=p_2=constant=\frac{e^{-2c_0}}{4}$ and from Eq.(\ref{b:20}),
we obtain $q=0$. This case leads to the perfect fluid case and we
get the same solution as given in \cite{KSPSS21}. For $\rho_1=0$,
Eq.(\ref{b:21}), we solve set of ODEs either for $\dot{\mu}=0$ or
$\dot{\mu}=1$. The case $\dot{\mu}=0$ provides the same solution as
we can obtain for the earlier case $\dot{\mu}=0$, which leads to the
perfect fluid case \cite{KSPSS21}. When $\dot{\mu}=1$,
Eq.(\ref{b:21}) implies that either $\dot{\lambda}=-1$ or
$\dot{\lambda}=-3$. Both the possibilities do not provide solution.
Also, the \textbf{Case IV} gives the same behavior as the
\textbf{Case II}.

\subsubsection*{EOS (3)}

When a perfect fluid satisfies EOS(3), Eqs.(\ref{b:10}) and
(\ref{b:12}) yield
\begin{equation}\label{b:32}
p_1=k\rho_1, \quad p_2=k\rho_2, \quad [Case~V]
\end{equation}
where $-1\leq k\leq1$, $k\neq0$. We explore the solutions either for
$k=-1$ or $k\neq-1$. When $k=-1$, basic equations contradict. For
$k\neq-1$, the case $\rho_1=0=\rho_2$ provides $q=0$ which reduces
to the perfect fluid case. The case, when
$\rho_1=0,~\rho_2=arbitrary$, Eq.(\ref{b:21}) implies that either
$1+\dot{\lambda}+2\dot{\mu}=0$ or $\dot{\lambda}=-1$. The first
option gives no solution while for $\dot{\lambda}=-1$,
Eqs.(\ref{b:22})-(\ref{b:24}) imply that $\dot{\mu}=0$ and
$\alpha=-2$ then Eq.(\ref{b:20}) provides $q=0$. This leads to the
perfect fluid case \cite{KSPSS21} and gives the same solution as we
have for EOS(2). For $\rho_2=0,~\rho_1=arbitrary$, Eq.(\ref{b:19})
implies that either $\dot{\lambda}=0$ or $\dot{\lambda}=-2\dot{\mu}$
both yield contradiction. The case $\rho_1,~\rho_2=arbitrary$ gives
contradiction in the basic equations. Also, for $k=1$, we have no
solution.

\subsubsection{Self-similarity of the Zeroth Kind}

The EFEs show that the quantities $\rho,~p$ and $q$ must be of the
form
\begin{eqnarray}\label{b:36}
\kappa\rho &=& \frac{1}{x^2}[\rho_1(\xi)+x^2\rho_2(\xi)],\\
\label{b:37}\kappa q&=&\frac{1}{x}{q(\xi)},\\
\label{b:38}\kappa p &=&\frac{1}{x^2}[p_1(\xi)+x^2p_2(\xi)],
\end{eqnarray}
where the self-similar variable is $\xi=\frac{x}{e^{t}}$. A set of
ODEs yield
\begin{eqnarray}
\label{pr:85}\dot{\rho_1}&=&-(\dot{\mu}+2\dot{\lambda})(\rho_1+p_1)
+{e^\mu}(\dot{q}+q+3q\dot{\mu}+2q\dot{\lambda}),\\
\label{pr:86} \dot{\rho_2}&=&-(\dot{\mu}+2\dot{\lambda})(\rho_2+p_2),\\
\label{pr:87}-\dot{p_1}+2p_1 &=& \dot{\mu}(\rho_1+p_1),\\
\label{pr:88}-\dot{p_2}
&=&\dot{\mu}(\rho_2+p_2)+e^{\mu}(-\dot{q}-2q\dot{\lambda}-3q\dot{\mu}),\\
\label{pr:89}{e^{2\mu}}\rho_1&=&-4\dot{\lambda}-3{\dot{\lambda}}^{2}-2\ddot{\lambda}-1
+2\dot{\mu}+2\dot{\lambda}\dot{\mu},\\
\label{pr:90}e^{2\mu}\rho_2 &=&{\dot{\lambda}}^{2}+2\dot{\lambda}\dot{\mu},\\
\label{pr:91}e^{3\mu}q&=&2(-\ddot{\lambda}-{\dot{\lambda}}^2
-\dot{\lambda}+2\dot{\lambda}\dot{\mu}+\dot{\mu}),\\
\label{pr:92}{e^{2\mu}}p_1&=&1+2\dot{\lambda}+{\dot{\lambda}}^2
+2\dot{\mu}+2\dot{\lambda}\dot{\mu},\\
\label{pr:93}e^{2\mu}p_2&=&2\dot{\lambda}\dot{\mu}
-2\ddot{\lambda}-3{\dot{\lambda}}^2,\\
\label{pr:94}e^{2\mu}p_1&=&\ddot{\lambda}+{\dot{\lambda}}^2
+\dot{\lambda}+\ddot{\mu}-\dot{\mu},\\
\label{pr:95}e^{2\mu}p_2&=&-\ddot{\lambda}-{\dot{\lambda}}^2-\ddot{\mu}.
\end{eqnarray}
Proceeding in a similar way as we have done for the first and
second kinds, we obtain either contradiction or $q=0$. This
reduces to the perfect fluid case already given in the literature
\cite{KSPSS21}.

\subsection{Tilted Dust, Orthogonal Fluid and Dust and Parallel Dust Cases}

For the dust case, we take $p=0$ in the equations of the fluid
case. Proceeding in a similar fashion as above, we ultimately
arrive either at contradiction or $q=0$ and hence reduces to the
perfect fluid and dust cases \cite{KSPSS21}.

\subsection{Parallel Fluid Case}

\subsubsection{Self-similarity of the Second Kind}

In this kind, the EFEs indicate that the quantities $\rho$, $p$
and $q$ must be of the following form
\begin{eqnarray}
\label{E:PdSSP2}\kappa\rho &=&t^{-2}\rho_1(\xi)+t^{-4}\rho_2(\xi),\\
\kappa q &=&t^{-4}q(\xi),\\
\label{E:PpSSP2}\kappa p &=& t^{-2}p_1(\xi)+t^{-4}p_2(\xi).
\end{eqnarray}
For this kind, a set of ODEs will be
\begin{eqnarray}\label{E:PSEP21}
e^{3\mu}q&=&-2\mu',\\\label{E:PSEP22} e^{2\mu}\rho_1&=&
2\lambda'\mu'-3{\lambda'}^2-2\lambda'',\\\label{E:PSEP23}
e^{2\mu}\rho_2&=&3,\\\label{E:PSEP24} e^{2\mu}p_1 &=&
{\lambda'}^2+2\lambda'\mu',\\\label{E:PSEP25} e^{2\mu}p_2
&=&1,\\\label{E:PSEP26} e^{2\mu}p_1&=&
\lambda''+{\lambda'}^2+\mu'',\\\label{E:PSEP28}
\rho_1+3p_1&=&-e^{\mu}(q'+3q\mu'+2q\lambda'),\\\label{E:PSEP29}
3p_2&=&\rho_2,\\\label{E:PSEP210}
-p'_1&=&\mu'(\rho_1+p_1),\\\label{E:PSEP211}
-p'_2&=&\mu'(\rho_2+p_2)+e^{\mu}q,
\end{eqnarray}
where prime indicates derivative with respect to $\xi=x$.

\subsubsection*{EOS(1) and EOS(2)}

When a perfect fluid satisfies EOS(1), Eqs.(\ref{E:PdSSP2}) and
(\ref{E:PpSSP2}) give that
\begin{eqnarray}\label{E:PEOS12P1}
p_2=0=\rho_1,\quad \alpha=2,\quad \gamma=\frac{1}{2},\quad p_1=
k(8\pi G)^{(1-\gamma)}{\rho_2}^\gamma.\quad [Case~I]
\end{eqnarray}
For EOS(2), it turns out that
\begin{eqnarray}\label{E:PEOS22P2}
p_2=0,~ \alpha=2,~ \gamma=\frac{1}{2},~
p_1=\frac{k}{{m_b}^{\gamma} (8\pi
G)^{(\gamma-1)}}{\rho_2}^\gamma=(\gamma-1)\rho_1.~ [Case~II]
\end{eqnarray}
For both cases, Eq.(\ref{E:PSEP25}) gives a contradiction.

\subsubsection*{EOS(3)}

For EOS(3), Eqs.(\ref{E:PdSSP2}) and (\ref{E:PpSSP2}) show that
\begin{equation}\label{E:PEOS32P3}
p_1=k\rho_1,\quad  p_2=k\rho_2.\quad [Case~III]
\end{equation}
Here Eq.(\ref{E:PSEP29}) implies that $k=\frac{1}{3}$. Equations
(\ref{E:PSEP22}), (\ref{E:PSEP24}) and (\ref{E:PSEP26}) provide a
relation $\mu''+3{\lambda'}^2+2\lambda''=0$. Now by taking
$\lambda=const$, we get $\mu''=0$. Since $\mu=\mu(\xi)$,
Eq.(\ref{E:PSEP21}) gives $q=-2e^{-3\mu}\mu'$. Eqs.(\ref{E:PSEP22})
and (\ref{E:PSEP23}) yield $\rho_1=0=p_1$ and $p_2=e^{-2\mu}$
respectively. Finally, we arrive at the following solution
\begin{eqnarray}
\lambda&=&constant,\quad \mu=\mu(\xi), \quad
\rho_1=0=p_1,\nonumber\\\label{E:PPP2S2} \quad \rho_2&=&3e^{-2\mu},
\quad p_2=e^{-2\mu}, \quad k=\frac{1}{3}.
\end{eqnarray}
This gives the following spacetime
\begin{equation}\label{E:PPP2S2CM}
ds^2=t^2e^{2\mu(x)}(dt^2-dx^2)-t^2(dy^2+ dz^2).
\end{equation}

\subsubsection{Self-similarity of the Infinite Kind}

A set of ODEs is given by
\begin{eqnarray} \label{E:PSEPinf1}
e^{2\mu}\rho&=&2\lambda'\mu'-3{\lambda'}^2-2\lambda'',\\\label{E:PSEPinf2}
e^{2\mu}p &=&{\lambda'}^2+2\lambda'\mu',\\\label{E:PSEPinf3}
e^{2\mu}p&=&\lambda''+{\lambda'}^2+\mu'',\\\label{E:PSEPinf4}
-p'&=&\mu'(\rho+p),\\\label{E:PSEPinf5}q&=&0.
\end{eqnarray}
Equation (\ref{E:PSEPinf5}) implies that $q=0$ which reduces to the
perfect fluid case and provides the same solution as in
\cite{KSPSS21}.

\section{Kinematic Self-similar Charge Dust Solutions}

Here we find KSS charge solutions of the plane symmetric spacetime
given by Eq.(\ref{a:2}). We restrict here to explore solutions for
the dust case only. The energy-momentum tensor for the sum of dust
and electromagnetic field can be written as
\begin{equation}\label{A:2}
T_{ab}=\rho{u_a}{u_b}-\frac{1}{4{\pi}}(F_{ac}F_{b}^{~c}-\frac{1}{4}g_{ab}F_{cd}F^{cd}).
\end{equation}
The Maxwell's field tensor $F_{ab}$ is defined as
\begin{equation}\label{A:3}
F_{ab}={\phi}_{b,a}-{\phi}_{a,b},
\end{equation}
where $\phi_a$ is the four-potential. Since charge is assumed to
be at rest with respect to the co-moving coordinate system, there
is no magnetic field present in this system \cite{ZAK MAX}. Thus
we can write $\phi_a=({\phi}(t,x),0,0,0)$. We would like to
mention here that ${\phi}(t,x)$ and $J$ are related by the
Maxwell's equation $F^{ab}_{;{b}}=-4\pi J^a$, where $F^{ab}$ is
the Maxwell field tensor which involves potential. In order to
find solution of the EFEs, we need to calculate the non-zero
components of $T_{ab}$ which can be obtained by $F_{ab}$. The only
non-zero components of the field tensor are \cite{KSPSS25b}
$-F_{01}=F_{10}=\phi_x$ which can also be written in contravariant
form as $F^{01}=-F^{10}=\phi_x e^{-4\mu}$. Thus the EFEs become
\begin{eqnarray}
\label{A:7}\kappa\rho&=&e^{-2\mu}(-3{\lambda_{x}}^{2}-2\lambda_{xx}
+{\lambda_{t}}^{2}+2{\lambda_{x}}{\mu_{x}}+2{\lambda_{t}}{\mu_{t}})-
\frac{\kappa}{8\pi}{\phi_x}^2{e^{-4\mu}},\\
\label{A:8}0&=&\lambda_{tx}-\lambda_{t}\mu_{x}+\lambda_{t}\lambda_{x}-\lambda_{x}\mu_{t},\\
\label{A:9}0&=&
e^{-2\mu}({\lambda_{x}}^{2}+2{\lambda_{x}}{\mu_{x}}-2\lambda_{tt}-3{\lambda_{t}}^{2}+2\lambda_{t}\mu_{t})
+\frac{\kappa}{8\pi}{\phi_x}^2{e^{-4\mu}},\\
\label{A:10}0&=&
e^{-2\mu}({\lambda_{x}}^{2}-\lambda_{tt}-{\lambda_{t}}^{2}+\lambda_{xx}+\mu_{xx}-\mu_{tt})
-\frac{\kappa}{8\pi}{\phi_x}^2{e^{-4\mu}}.
\end{eqnarray}
The energy-momentum conservation provides
\begin{eqnarray}\label{A:11}
\mu_{t}&=&-\frac{\rho_t}{\rho}-2\lambda_{t}-
\frac{e^{-4\mu}\phi_x}{4\pi\rho}(-2\phi_x\mu_t+\phi_{tx}+2\phi_x\lambda_t),\\
\label{A:12} \mu_{x}&=&-
\frac{e^{-4\mu}\phi_x}{4\pi\rho}(2\phi_x\mu_x-\phi_{xx}-2\phi_x\lambda_x).
\end{eqnarray}

\subsection{Tilted Dust Case}

\subsubsection{Self-similarity of the First Kind}

We can take $\phi(t,x)=\phi(\xi)$, where the self-similar variable
is $\xi=\frac{x}{t}$. It follows from the EFEs that energy density
$\rho$ must be of the form given in Eq.(\ref{b:2}). Then from the
field equations and the equations of motion, i.e.,
Eqs.(\ref{A:7})-(\ref{A:12}), we obtain the following set of ODEs
\begin{eqnarray}
\label{T1:i}\dot{\rho}&=&-(\dot{\mu}+2\dot{\lambda})\rho
+\frac{\kappa}{4\pi}e^{-4\mu(\xi)}\dot{\phi}(2\dot{\phi}\dot{\mu}
-\ddot{\phi}-2\dot{\phi}\dot{\lambda}),\\
\label{T1:ii}\dot{\mu}\rho&=&-\frac{\kappa}{4\pi}
e^{-4\mu(\xi)}\dot{\phi}(2\dot{\phi}\dot{\mu}
-\ddot{\phi}-\dot{\phi}-2\dot{\phi}\dot{\lambda}),\\
\label{T1:iii}\rho &=& e^{-2\mu(\xi)}(-4\dot{\lambda}
-3{\dot{\lambda}}^{2}-2\ddot{\lambda}
-1+2\dot{\mu}+2\dot{\lambda}\dot{\mu})-\frac{\kappa}{8\pi}e^{-4\mu(\xi)}\dot{\phi}^2,\\
\label{T1:iv}0&=&{\dot{\lambda}}^{2}+2\dot{\lambda}\dot{\mu},\\
\label{T1:v}0&=&-\ddot{\lambda}-{\dot{\lambda}}^2
-\dot{\lambda}+2\dot{\lambda}\dot{\mu}+\dot{\mu},\\
\label{T1:vi}0&=&1+2\dot{\lambda}+{\dot{\lambda}}^2+2\dot{\mu}
+2\dot{\lambda}\dot{\mu}+\frac{\kappa}{8\pi}e^{-2\mu(\xi)}\dot{\phi}^2,\\
\label{T1:vii}0&=&-2\dot{\lambda}\dot{\mu}+2\ddot{\lambda}
+3{\dot{\lambda}}^2+2\dot{\lambda},\\
\label{T1:viii}0&=&\ddot{\lambda}+{\dot{\lambda}}^2+\dot{\lambda}
-\dot{\mu}+\ddot{\mu}
-\frac{\kappa}{8\pi}e^{-2\mu(\xi)}\dot{\phi}^2,\\
\label{T1:ix}0&=&\ddot{\lambda}+{\dot{\lambda}}^2
+\dot{\lambda}+\dot{\mu}+\ddot{\mu}.
\end{eqnarray}
The above set of equations yields contradiction.

\subsubsection{Self-similarity of the Second Kind}

The self-similar variable for this kind is $\xi=\frac{x}{(\alpha
t)^\frac{1}{\alpha}}$ and hence $\phi=\phi(\xi)$. Thus the EFEs
and equations of motion yield
\begin{eqnarray}
\label{T2:i}\dot{\rho_1}
&=&-(\dot{\mu}+2\dot{\lambda})\rho_1+\frac{\kappa}{4\pi}e^{-4\mu(\xi)}
\dot{\phi}(2\dot{\phi}\dot{\mu}-\ddot{\phi}-2\dot{\phi}\dot{\lambda}),\\
\label{T2:ii}\dot{\rho_2}+2\alpha\rho_2&=&-(\dot{\mu}+2\dot{\lambda})\rho_2,\\
\label{T2:iii}\dot{\mu}\rho_1&=&-\frac{\kappa}{4\pi}e^{-4\mu(\xi)}
\dot{\phi}(2\dot{\phi}\dot{\mu}-\ddot{\phi}-\dot{\phi}-2\dot{\phi}\dot{\lambda}),\\
\label{T2:iv}0&=&\dot{\mu}\rho_2,\\
\label{T2:v}\rho_1 &=&e^{-2\mu(\xi)}(
-4\dot{\lambda}-3{\dot{\lambda}}^{2}
-2\ddot{\lambda}-1+2\dot{\mu}+2\dot{\lambda}\dot{\mu})\nonumber\\&&
-\frac{\kappa}{8\pi}e^{-4\mu(\xi)}\dot{\phi}^2,\\
\label{T2:vi}\alpha^2e^{2\mu}\rho_2 &=&{\dot{\lambda}}^{2}+2\dot{\lambda}\dot{\mu},\\
\label{T2:vii}0&=&-\ddot{\lambda}
-{\dot{\lambda}}^2-\dot{\lambda}+2\dot{\lambda}\dot{\mu}+\dot{\mu},\\
\label{T2:viii}0&=&1+2\dot{\lambda}+{\dot{\lambda}}^2
+2\dot{\mu}+2\dot{\lambda}\dot{\mu}+\frac{\kappa}{8\pi}e^{-2\mu(\xi)}\dot{\phi}^2,\\
\label{T2:ix}0 &=&2\ddot{\lambda}+3{\dot{\lambda}}^2
+2\alpha\dot{\lambda}-2\dot{\lambda}\dot{\mu},\\
\label{T2:x}0&=&\ddot{\lambda}+{\dot{\lambda}}^2
+\dot{\lambda}-\dot{\mu}+\ddot{\mu}-\frac{\kappa}{8\pi}e^{-2\mu(\xi)}\dot{\phi}^2,\\
\label{T2:xi}0&=&\ddot{\lambda}+{\dot{\lambda}}^2
+\alpha\dot{\lambda}+\alpha\dot{\mu}+\ddot{\mu}.
\end{eqnarray}
Equation (\ref{T2:iv}) implies that either $\dot{\mu}=0$ or
$\rho_2=0$. The first possibility, $\dot{\mu}=0$, yields
contradiction while for the second possibility $\rho_2=0$,
Eq.(\ref{T2:vi}) gives either $\dot{\lambda}=0$ or
$\dot{\lambda}=-2\dot{\mu}$, both yield contradiction.

\subsubsection{Self-similarity of the Zeroth Kind}

For $\phi=\phi(\xi)$, where $\xi=\frac{x}{e^{t}}$,
Eqs.(\ref{A:7})-(\ref{A:12}) yield the following set of ODEs
\begin{eqnarray}
\label{T0:i}\dot{\rho_1}&=&-(\dot{\mu}+2\dot{\lambda})\rho_1
+\frac{\kappa}{4\pi}e^{-4\mu(\xi)}\dot{\phi}(2\dot{\phi}\dot{\mu}
-\ddot{\phi}-2\dot{\phi}\dot{\lambda}),\\
\label{T0:ii}\dot{\rho_2} &=&-(\dot{\mu}+2\dot{\lambda})\rho_2,\\
\label{T0:iii}\dot{\mu}\rho_1&=&-\frac{\kappa}{4\pi}
e^{-4\mu(\xi)}\dot{\phi}(2\dot{\phi}\dot{\mu}
-\ddot{\phi}-\dot{\phi}-2\dot{\phi}\dot{\lambda}),\\
\label{T0:iv}0&=&\dot{\mu}\rho_2,\\
\label{T0:v}\rho_1&=&e^{-2\mu(\xi)}(-4\dot{\lambda}-3{\dot{\lambda}}^{2}
-2\ddot{\lambda}-1+2\dot{\mu}+2\dot{\lambda}\dot{\mu})-\frac{\kappa}{8\pi}e^{-4\mu(\xi)}\dot{\phi}^2,\\
\label{T0:vi}e^{2\mu}\rho_2&=&{\dot{\lambda}}^{2}+2\dot{\lambda}\dot{\mu},\\
\label{T0:vii}0&=&\ddot{\lambda}
+{\dot{\lambda}}^2+\dot{\lambda}-2\dot{\lambda}\dot{\mu}-\dot{\mu},\\
\label{T0:viii}0&=&1+2\dot{\lambda}
+{\dot{\lambda}}^2+2\dot{\mu}+2\dot{\lambda}\dot{\mu}
+\frac{\kappa}{8\pi}e^{-2\mu(\xi)}\dot{\phi}^2,\\
\label{T0:ix}0&=&-2\dot{\lambda}\dot{\mu}
+2\ddot{\lambda}+3{\dot{\lambda}}^2,\\
\label{T0:x}0&=&\ddot{\lambda}+{\dot{\lambda}}^2
+\dot{\lambda}-\dot{\mu}+\ddot{\mu}-\frac{\kappa}{8\pi}e^{-2\mu(\xi)}\dot{\phi}^2,\\
\label{T0:xi}0&=&-\ddot{\lambda}-{\dot{\lambda}}^2-\ddot{\mu}.
\end{eqnarray}
Equation (\ref{T0:iv}) implies that either $\dot{\mu}=0$ or
$\rho_2=0$, both give contradiction.

\subsection{Orthogonal Dust Cases}

In this case, set of basic equations contradict for the
self-similarity of the zeroth kind.

\subsection{Parallel Dust Case}

\subsubsection{Self-similarity of the Second Kind}

For this kind, we have $\xi=x,~\phi(t,x)=\phi(\xi)$ and $\rho$ given
in Eq.(\ref{E:PdSSP2}). Thus a set of ODEs will be
\begin{eqnarray}\label{F:PSEP21}
\mu'&=&0,\\\label{F:PSEP22} e^{2\mu}\rho_1&=&
2\lambda'\mu'-3{\lambda'}^2-2\lambda'',\\\label{F:PSEP23}
e^{2\mu}\rho_2&=&3-\frac{\kappa}{8\pi}e^{-2\mu(\xi)}{\phi^\prime}^2,\\\label{F:PSEP24}
0&=&2\lambda'\mu'+{\lambda'}^2,\\\label{F:PSEP25}
1 &=&-\frac{\kappa}{8\pi}e^{-2\mu(\xi)}{\phi^\prime}^2,\\
\label{F:PSEP26}0&=&
\lambda''+{\lambda'}^2+\mu'',\\\label{F:PSEP27}
1&=&\frac{\kappa}{8\pi}e^{-2\mu(\xi)}{\phi^\prime}^2,\\\label{G:PSEP29}\rho_1&=&0,\\
\label{H:PSEP29}\rho_2&=&0,\\\label{F:PSEP28}
{\mu^\prime}{\rho_2}&=&-\frac{\kappa}{4\pi}e^{-4\mu(\xi)}{\phi^\prime}
(2{\phi^\prime}{\mu^\prime}-{\phi^{\prime\prime}}-2{\phi^\prime}{\lambda^\prime}),\\\label{F:PSEP29}
{\mu^\prime}{\rho_1}&=&0,
\end{eqnarray}
Eqs.(\ref{F:PSEP25}) and (\ref{F:PSEP27}) clearly give
contradiction.

\subsubsection{Self-similarity of the Infinite Kind}

A set of ODEs in terms of $\xi=x$ is
\begin{eqnarray} \label{E:PSEPinf1}
e^{2\mu}\rho&=&2\lambda'\mu'-3{\lambda'}^2-2\lambda''-\frac{\kappa}{8\pi}{\phi'}^{2}e^{-2\mu},\\
\label{E:PSEPinf2}0&=&{\lambda'}^2+2\lambda'\mu'+\frac{\kappa}{8\pi}{\phi'}^{2}e^{-2\mu},\\
\label{E:PSEPinf3}0&=&{\lambda'}^2+\lambda''+\mu''-\frac{\kappa}{8\pi}{\phi'}^{2}e^{-2\mu},\\
\label{E:PSEPinf4}e^{2\mu}\mu'\rho&=&-\frac{\kappa}{4\pi}{\phi'}e^{-2\mu}(2\phi'\mu'-\phi''-2\lambda'\phi').
\end{eqnarray}
Equations (\ref{E:PSEPinf1}) and (\ref{E:PSEPinf2}) yield
\begin{equation}\label{A234}
\rho=2e^{-2\mu}(2\lambda'\mu'-\lambda''-{\lambda'}^2),
\end{equation}
Now Eqs.(\ref{E:PSEPinf2}) and (\ref{E:PSEPinf3}) imply
\begin{equation}\label{A567}
2\lambda'\mu'+2{\lambda'}^2+\mu''+\lambda''=0.
\end{equation}
Assume that $\mu=constant$ so that Eq.(\ref{A567}) gives
$\lambda''=-2{\lambda'}^2$ then Eq.(\ref{A234}) provides
$\rho=2e^{-2\mu}{\lambda'}^2$. Equation (\ref{E:PSEPinf4}) can be
written as $\phi'(-\phi''-2\lambda'\phi')=0$, which yields either
$\phi'=0$ or $\phi''=-2\lambda'\phi'$. For $\phi'=0$, we get
contradiction while for the second possibility
$\phi''=-2\lambda'\phi'$, we obtain a solution of the following form
\begin{eqnarray}\label{A890}
\lambda&=&\frac{1}{2}\ln c_1(2\xi-c),\quad
\mu=constant,\quad\rho=2e^{-2\mu}{\lambda'}^2,\nonumber\\
\phi&=&\frac{c_2}{2}\ln c_3(2\xi-c),\quad
\frac{\kappa}{8\pi}e^{-2\mu}=-\frac{1}{{c_2}^2}.
\end{eqnarray}
The corresponding metric is
\begin{equation}\label{B012}
ds^2=dt^2-dx^2-2x(dy^2+dz^2).
\end{equation}

\section{Outlook}

This paper is devoted to study the effects of heat conduction and
electromagnetic field on plane symmetric KSS solutions of the EFEs.
We have found these solutions by using heat conducting and charge
dust fluids. These types of fluid are real and exist in nature
rather than imaginary and ideal as perfect fluid. We have
investigated KSS solutions for the case when the KSS vector is
tilted, orthogonal and parallel to the fluid flow with either
EOS(1), EOS(2) or EOS(3).

Firstly, we have analyzed plane symmetric KSS heat conducting fluid
solutions. The self-similar variables and the metric functions
remain the same as for the perfect fluid case \cite{KSPSS21}. We
find \emph{only} one solution with non-zero heat flux $q$ given by
Eq.(\ref{E:PPP2S2}). The remaining cases either give contradiction
or reduce to the perfect fluid solutions already discussed in
\cite{KSPSS21}. The heat flux turns to be an arbitrary function with
negative sign which indicates dissipation. Secondly, we have
obtained \emph{one} plane symmetric KSS charge dust solution in the
parallel dust case with self-similarity of the infinite kind given
by Eq.(\ref{A890}). Here we obtain potential in terms of $x$ and its
sign depends upon the arbitrary constants. We would like to mention
here that energy density for both the solutions is positive.

It would be interesting to extend this analysis to find the KSS
charge perfect solution and also for the most general plane
symmetric spacetime.

\end{document}